\RequirePackage{fix-cm}
\documentclass[twocolumn]{svjour3}          
\usepackage[utf8]{inputenc}
\smartqed  
\usepackage{amssymb}
\usepackage{latexsym}
\usepackage{graphicx}
\begin{document}

\title{
$XY$-VBS phase boundary for the 
square-lattice 
$J_1$-$J_2$ $XXZ$ 
model with the ring exchange
}
\subtitle{}


\author{Yoshihiro Nishiyama
}

\institute{Department of Physics, Faculty of Science,
Okayama University, Okayama 700-8530, Japan }

\date{Received: date / Accepted: date}

\maketitle

e-mail\footnote{e-mail: nisiyama@science.okayama-u.ac.jp (corresponding author)}

\begin{abstract}

The 
	square-lattice 
	$J_1$-$J_2$ $XXZ$ 
	model with the ring-exchange 
interaction $K$
was investigated numerically.
As for 
the hard-core-boson model with
	the nearest-neighbor hopping $J_1/2$,
namely, the $J_1$-$K$ $XY$ model,
it has been reported that
the ring exchange leads to a variety of exotic phases such as the
valence-bond-solid (VBS) phase.
In this paper,
we
extend the parameter space in order to
investigate 
the phase boundary 
between the $XY$ (superfluid) and VBS phases.
A notable feature is that the phase boundary 
terminates at the fully-frustrated
point, $J_2/J_1 \to 0.5^-$.
As a scaling parameter for the multi-criticality,
the distance from the multi-critical point $\delta (\ge 0)$ is introduced.
In order to detect the phase transition,
we employed
the high-order 
fidelity susceptibility $\chi^{(3)}_F$,
which is readily evaluated via the exact-diagonalization scheme.
As a demonstration, for a fixed value of $\delta$,
the $XY$-VBS criticality was analyzed by 
the probe $\chi^{(3)}_F$. 
Thereby, with properly scaling $\delta$,
the $\chi^{(3)}_F$ data were cast into the crossover-scaling formula
to determine the multi-criticality.

\end{abstract}

\section{\label{section1}Introduction}

The two-dimensional hard-core boson model with the nearest-neighbor-hopping
amplitude
$J_1/2$
and the ring exchange around each plaquette $K$
has been investigated extensively \cite{Huerga14,Sandvik02}.
It has been shown that the ring exchange leads to
a variety of exotic phases such as the valence-bond-solid (VBS) phase.
The model is equivalent to 
the spin-$S=1/2$ $J$-$K$ $XY$ magnet \cite{Huerga14}.
Meanwhile,
the $J_1$-$J_2$ $XXZ$ model was investigated with the coupled cluster method
to surmount the negative sign problem \cite{Bishop08};
it has to be mentioned that a limiting case, namely,
the $J_1$-$J_2$ $XY$ model, has been studied with the tensor-network \cite{Chan12}
and density-matrix-renormalization-group \cite{Chan23} methods in depth.
The $J_1$-$J_2$ $XXZ$ model is equivalent to the hard-core-boson model
with the ``kinetic frustration'' $J_2$ \cite{Chen17}, and the first- and second-neighbor Coulomb repulsions.
The VBS phase is stabilized by the kinetic frustration
as well as the Coulomb repulsions.
A notable feature is that
the phase boundary between the $XY$ (superfluid) and VBS phases
terminates at the fully frustrated point $J_2/J_1=0.5$ eventually.
Such a $J_2$-mediated multi-criticality poses an intriguing problem.

In this paper,
we investigate the $J_1$-$J_2$-$K$ $XXZ$ model 
by means of the
exact-diagonalization method.
Our
main concern is to investigate the $XY$-VBS phase boundary
from the viewpoint of the extended parameter space.
In order to detect the $XY$-VBS phase transition,
we employ
the ``high order'' \cite{Wang09,Lv22} fidelity susceptibility \cite{Quan06,Zanardi06,HQZhou08,Yu09,You11}, which is readily 
evaluated
via the exact diagonalization scheme.
The high-order fidelity susceptibility is also applied to the analysis of the multi-criticality
toward $J_2/J_1 \to 0.5^-$.

To be specific, we present the Hamiltonian for the $J_1$-$J_2$-$K$ $XXZ$ model
\begin{eqnarray}
\nonumber
{\cal H} &=&
- J_1\sum_{\langle i j\rangle} (S^x_iS^x_j + S^y_iS^y_j) 
+J_1\Delta\sum_{\langle ij\rangle} S^z_iS^z_j  
+ J_2\sum_{\langle\langle ij\rangle\rangle}(S^x_iS^x_j + S^y_iS^y_j) \\ 
& & + J_2\Delta\sum_{\langle \langle ij\rangle\rangle} S^z_iS^z_j
+K\sum_{[ijkl]}  (S^+_iS^+_kS^-_jS^-_l+
S^+_lS^+_jS^-_kS^-_i) 
.
\label{Hamiltonian}
\end{eqnarray}
Here,
the quantum spin-$1/2$ operators $\{ {\bf S}_i\}$ are placed at each square-lattice point, $i=1,2,\dots,N$
($N$: total number of spins),
and the symbols $\{ S^\pm_i \}$ denote the ladder operators.
The summations,
$\sum_{\langle ij \rangle}$ and
$\sum_{\langle\langle ij \rangle\rangle}$, run over all possible nearest- and next-nearest-neighbor
pairs, 
$\langle ij\rangle$ and $\langle \langle ij \rangle\rangle$, respectively.
Similarly, the summation $\sum_{[ijkl]}$ runs over all possible plaquette spins,
$[ijkl]$;
here, the indices are
arranged like $^i_l\Box^j_k$ around each plaquette, $\Box$.
The parameters, $J_1$, $J_2$, and $K$,
denote the nearest-neighbor, next-nearest-neighbor, and ring exchange interactions, respectively,
and the anisotropy parameter is given by $\Delta$.
Hereafter, the nearest-neighbor interaction $J_1$ is regarded as the unit of
energy, {\it i.e.}, $J_1=1$.
In the hard-core-boson language \cite{Huerga14},
the parameters, $J_{1.2}/2$ and $J_{1,2} \Delta$, correspond the hopping amplitudes and inter-site Coulomb repulsions,
respectively.

\begin{figure}
	\includegraphics[width=90mm]{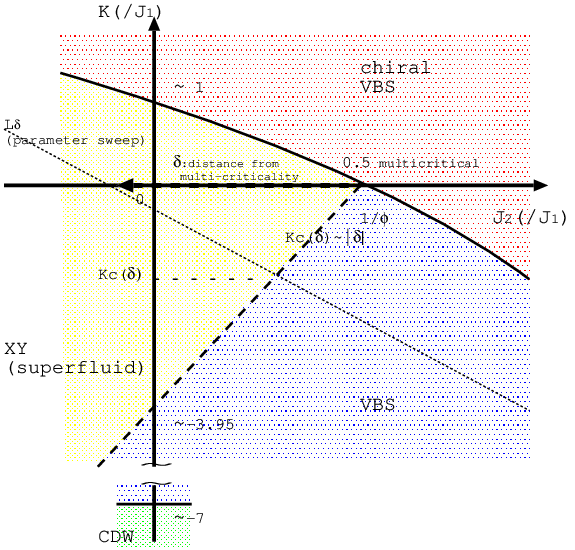}
\caption{\label{figure1}
A schematic phase diagram for the $J_1$-$J_2$-$K$ model 
(\ref{Hamiltonian})
is shown. Here, the anisotropy
is set to $\Delta=-0.15$ so as to locate the multi-critical point 
	$(0.5,0)$ within the
abscissa axis \cite{Bishop08}.
In this paper \cite{Bishop08},
the abscissa axis $K=0$  was investigated,
whereas the ordinate axis $J_2=0$ (albeit, for $\Delta=0$)
was studied in Ref. \cite{Huerga14,Sandvik02}.
	The $XY$ (superfluid), valence bond solid (VBS), chiral VBS, and charge density wave (CDW) phases appear successively,
as $J_2(/J_1)$ and $K(/J_1)$ change.
The dashed (solid) line indicates that the phase transition is
(dis)continuous \cite{Huerga14,Sandvik02,Bishop08}, and this continuous branch
is our main concern.
The parameters, $J_2$ and $K$, are swept along the line
L$_\delta$ (\ref{parameter_space}) parameterized by
$\delta$
(distance from the multi-criticality).
The line L$_\delta$ is 
almost parallel to the phase boundary
between the VBS and chiral-VBS phases so as to avoid the influence from this branch.
The crossover critical exponent $\phi$ is introduced in order to analyze the multi-criticality.
}
\end{figure}

A schematic phase diagram for the $J_1$-$J_2$-$K$ $XXZ$ model (\ref{Hamiltonian})
is presented in
Fig. \ref{figure1}.
Here, the anisotropy is set to $\Delta=-0.15$
so as to locate the multi-critical (quantum triple) point
\begin{equation}
\label{multi-critical_point}
(J_2,K)=(0.5,0)
,
\end{equation}
within the abscissa axis \cite{Bishop08}.
The $XY$ (superfluid), VBS, chiral VBS, and charge density wave (CDW) phases appear according to 
the preceeding studies, Ref. \cite{Bishop08} and \cite{Huerga14,Sandvik02};
in the former study \cite{Bishop08}, the case of 
the abscissa axis, $K=0$, was clarified,
whereas in the latter \cite{Huerga14,Sandvik02}, the ordinate axis $J_2=0$ (albeit, for $\Delta=0$) was investigated
in detail.
Here, we follow the terminology of Ref. \cite{Huerga14}, because this paper
deals with both $K>0$ and $K<0$ cases.
The dashed (solid) line indicates that 
the phase transition is
(dis)continuous \cite{Huerga14,Sandvik02,Bishop08}.
The interaction parameters, $J_2$ and $K$, are swept along the line
\begin{equation}
\label{parameter_space}
{\rm L}_\delta: \ 
K=-6(J_2-0.5+\delta)
,
\end{equation}
parameterized by $\delta$.
As explained in Fig. \ref{figure1},
the parameter $\delta$ indicates the distance from the multi-critical point
(\ref{multi-critical_point}), which plays a significant role in the crossover-scaling analysis.
In order to analyze the $XY$-VBS criticality reliably,
the slope of the line L$_\delta$ is set to be parallel to the phase boundary between the VBS and chiral VBS phases.
This technique
is a key ingredient of the present finite-size-scaling analyses,
where the extention of the parameter space $J_2$-$K$ is crucial.
The crossover critical exponent
$\phi$
characterizes the power-law singularity of the phase boundary around the multi-critical point
(\ref{multi-critical_point}) \cite{Riedel69,Pfeuty74}.

The ring-exchange term of the Hamiltonian (\ref{Hamiltonian})
is rewritten as
\begin{eqnarray}
\nonumber
& & 2K\sum_{[ijkl]} (S^x_iS^x_jS^x_kS^x_l+S^y_iS^y_jS^y_kS^y_l
+S^x_iS^x_jS^y_kS^y_l+S^y_iS^y_jS^x_kS^x_l  \\
& & +S^y_iS^x_jS^x_kS^y_l
+S^x_iS^y_jS^y_kS^x_l-S^x_iS^y_jS^x_kS^y_l-S^y_iS^x_jS^y_kS^x_l)
,
\end{eqnarray}
in terms of the ordinary $S_i^{x,y}$-based representation \cite{Huerga14}.
It is a good position to introduce
the $J$-$Q$ model \cite{Sandvik07,Melko08}.
The plaquette interaction term of the ``easy plane'' \cite{Emidio16} 
$J$-$Q_\perp$ model admits a neat expression
\begin{equation}
-Q_\perp \sum_{[ijkl]}(B_{ij}B_{kl}+B_{il}B_{jk})
,
\end{equation}
with the $XY$ interaction, $B_{ij}=2(S^x_iS^x_j+S^y_iS^y_j)$.
The plaquette interactions of the $J$-$K$ and $J$-$Q_\perp$ models differ significantly.
In fact, the $Q_\perp$ interaction
stabilizes the VBS phase,
whereas the exceedingly large $|K|$ stabilizes the CDW phase eventually, as shown in Fig. \ref{figure1}.
It is claimed that
the $Q_\perp$-driven phase transition to the VBS phase would be discontinuous \cite{Emidio16,Desai20}.

The rest of this paper is organized as follows.
In Sec. \ref{section2}, we present the numerical results.
The high-order fidelity susceptibility as well as
its scaling theory are explained.
In Sec. \ref{section3}, we address the summary and discussions.

\section{\label{section2}Numerical results}

In this section, we present the numerical results
for the $J_1$-$J_2$-$K$ $XXZ$ model (\ref{Hamiltonian}).
We employed
the exact diagonalization method
for the rectangular cluster with $N (=L^2) \le 6^2$ spins.
Here, the symbol $L$ denotes the linear dimension of the finite-size cluster.
The total magnetization 
$\sum_i^N S^z_i$
is fixed to
\begin{equation}
1 \ \ {\rm and }\ \ 1/2
,
\end{equation}
(magnon's dilution limit)
for the even- and odd-$N$ clusters, respectively,
so that 
these two series of data behave similarly;
even- and odd-$N$ systems show distinctive behaviors,
unless otherwise remedied.
Moreover, 
it has been claimed that
the VBS-N\'eel phase transition is detected sensitively by the inclusion of dilution
\cite{Poilblanc06}.

In order to analyze the $XY$-VBS phase transition,
we utilize the high-order fidelity susceptibility
\cite{Wang09,Lv22}
\begin{equation}
\label{fidelity_susceptibility}
\chi^{(3)}_F= -\partial_\lambda^3 F(\lambda)|_{\lambda=0}
.
\end{equation}
Here,
the fidelity $F$ is given by the overlap
$F(\lambda)=|\langle \lambda|0 \rangle|$
\cite{Uhlmann76,Jozsa94,Peres84,Gorin06}
between 
the ground states $|\lambda \rangle$ 
of proximate perturbation parameters, $\lambda$ and $(\lambda=)0$.
The perturbed Hamiltonian
${\cal H}+\lambda V $ 
is given by
\begin{equation}
\label{perturbation}
V=\sum_{\langle\langle ij \rangle\rangle}
{\cal P}^{t_0}_{ij}
,
\end{equation}
with the projection operator
${\cal P}^{t_0}_{ij}=|t_0\rangle_{ij} \ {}_{ij}\langle t_0|$
onto a triplet sector $|t_0\rangle_{ij}=\frac{|+\rangle_i|-\rangle_j+|-\rangle_i|+\rangle_j}{\sqrt{2}}$
with zero longitudinal magnetic moment;
here, the vector $  | \pm \rangle_{i} $ denotes the spin state at site $i$. 
Note that the exact diagonalization method yields the ground-state vector $|\lambda\rangle$
explicitly, and one is able to evaluate the fidelity $F(\lambda)=|\langle \lambda|0 \rangle|$
straightforwardly.
As would be apparent from the definition (\ref{perturbation}),
the perturbation field $V$ is sensitive to the onset of the in-plane magnetization,
namely, the $XY$ phase.
The perturbation field $V$ is magnetic-moment-direction insensitive,
and so, it is
not a symmetry-breaking field.
In a preliminary stage, we found that the singularities of the specific heat as well as the conventional
fidelity susceptibility
are hard to appreciate; possibly, the specific-heat critical exponent takes a negative value.
Therefore, we resort to
the high-order 
fidelity susceptibility
$\chi^{(3)}_F$ to detect the criticality sensitively
\cite{Wang09,Lv22}.


According to the scaling theory \cite{You11,Schwandt09,Albuquerque10},
the high-order fidelity susceptibility obeys
the scaling formula
\begin{equation}
\label{scaling_formula}
\chi_F^{(3)}=L^x f\left((K-K_c)L^{1/\nu}\right)
,
\end{equation}
with the critical point $K_c$, correlation-length critical exponent 
$\nu$,
$\chi^{(3)}_F$'s scaling dimension $x$,
and a scaling function $f$;
namely, the correlation length $\xi$ diverges as 
$\xi \sim |K-K_c|^{-\nu}$ at at the critical point.
(Note that the parameters $(J_2,K)$ are swept along the line L$_\delta$ (\ref{parameter_space}).)
The scaling dimension $x$ is given by
\begin{equation}
\label{scaling_dimension}
x=3/\nu
,
\end{equation}
because the operator $\partial_\lambda $ in Eq. (\ref{fidelity_susceptibility}) has the scaling dimension 
$1/\nu$, whereas the fidelity $F$ is dimensionless \cite{You11,Schwandt09,Albuquerque10}.
Hence, the high-order fidelity susceptibility has an enhanced scaling dimension.
One is able to capture the signature for the criticality sensitively,
even though the underlying singularity $1/\nu$ takes a small value.

\subsection{\label{section2_1}
$XY$-VBS phase transition with $\Delta=0$ and fixed $\delta$:
High-order-fidelity-susceptibility $\chi_F^{(3)}$ analysis}

In this section, we analyze the $XY$-VBS criticality via the probe $\chi^{(3)}_F$.
As for 
$J_2=0$ and $\Delta=0$, 
the stochastic-series-expansion  result 
\cite{Sandvik02}
is available. 
The transition point was estimated as 
$(J_{2c},K_c)=(0,-3.95)$ in our unit of energy, $J_1=1$.
In order to detect this transition point,
we set 
\begin{equation}
\label{delta_1.16}
\delta=  0.5 +3.95/6
.
\end{equation}
Then,
the parameter-sweep line L$_\delta$ (\ref{parameter_space})
passes the above-mentioned critical point $(0,-3.95)$.

\begin{figure}
	\includegraphics[width=90mm]{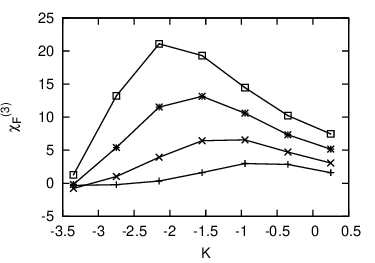}
\caption{\label{figure2}
	The high-order fidelity susceptibility $\chi^{(3)}_F$
	(\ref{fidelity_susceptibility})
	is plotted for various $K$, and 
	($+$) $L=3$,
	($\times$) $4$,
	($*$) $5$, and
	($\Box$) $6$,
	with $\delta=0.5+3.95/6$ 
	(\ref{delta_1.16})
	and $\Delta=0$.
	The signature for the $XY$-VBS phase transition
	is captured by $\chi_F^{(3)}$'s peak.
}
\end{figure}

In Fig. \ref{figure2},
we present the 
high-order fidelity susceptibility 
$\chi_F^{(3)}$
(\ref{fidelity_susceptibility})
for various $K$,
and 
($+$) $L=3$,
($\times$) $4$,
($*$) $5$, and
($\Box$) $6$
with $\delta=0.5+3.95/6$ (\ref{delta_1.16})
and $\Delta=0$.
The peak of $\chi_F^{(3)}$ develops rapidly, as the system size $L$ enlarges.
The peak indicates the onset of the $XY$-VBS phase transition at $K = K_c$.
For small $K < K_c$, the $XY$ phase is realized,
whereas for large $K>K_c$, the VBS phase emerges.
In contrast,
the background, namely, contribution of the non-singular part, appears to be
rather small.
Such a feature is  a merit of the high-order fidelity susceptibility, which
picks up the singular part sensitively out of the background.
Although
the finite-size drift of the peak position seems to be non-negligible,
the drift distance is governed by the scaling theory.
According to the scaling formula (\ref{scaling_formula}),
the drift distance $\delta K$ should scale as $\sim 1/L^{1/\nu}$,
and in a systematic manner,
the critical point can be appreciated as follows.

\begin{figure}
	\includegraphics[width=90mm]{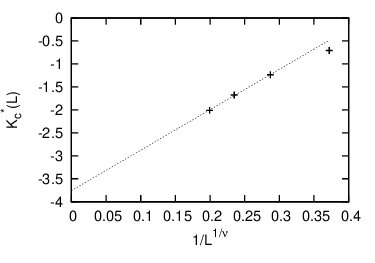}
\caption{\label{figure3}
The approximate critical point $K^*_c(L)$ (\ref{approximate_critical_point})
is plotted for $1/L^{1/\nu}$; the interaction parameters are the same as those
of Fig. \ref{figure2}.
The exponent is set to $1/\nu=0.9$.
The least-squares fit to the $L=4$-$6$ data yields an estimate
$K_c=-3.76(7)$ in the thermodynamic limit $L\to \infty$.
Possible systematic errors are considered in the text.
}
\end{figure}

In Fig. \ref{figure3},
we present the approximate critical point
$K_c^*(L)$ for $1/L^{1/\nu}$ with $\delta=0.5+3.95/6$ (\ref{delta_1.16}), $\Delta=0$
and $1/\nu=0.9$.
Here,
the approximate critical point denotes the position of the $\chi_F^{(3)}$ peak
\begin{equation}
\label{approximate_critical_point}
\frac{\partial \chi^{(3)}_F}{\partial K}|_{K=K_c^*(L)} =0
,
\end{equation}
for each $L$.
As explained above,
the abscissa scale $1/L^{1/\nu}$ describes the finite-size drift of $K_c^*(L)$,
and the validity of $1/\nu=0.9$ is considered afterward.
The least-squares fit to the $L=4$-$6$ data yields an estimate
$K_c=-3.76(7)$ in the thermodynamic limit $L\to\infty$.
In order to appreciate a possible systematic error,
replacing the abscissa scale $1/L^{1/\nu}$ with a smaller exponent $1/\nu=0.75$,
we made the similar analysis as that of Fig. \ref{figure3}.
(This lower bound $1/\nu=0.75$ is considered afterward.)
Thereby, we obtained an alternative value $K_c=-4.18(6)$.    
The deviation $\approx 0.42$ from the above estimate appears to
dominate the least-squares-fit error $\approx 0.07$,
and it may indicate a possible systematic error.
Therefore, considering the former as an indicator for the error margin
we obtained
\begin{equation}
\label{critical_point}
(J_{2c},K_c)=[
	-0.032(70),
	-3.76(42)
]
.
\end{equation}
The value of $J_{2c}$ comes from the propagation of uncertainty
through the relation (\ref{parameter_space}).
The present result (\ref{critical_point}) agrees with
the above-mentioned stochastic-series-expansion estimate
$(0,-3.95)$ \cite{Sandvik02},
indicating that the available simulation data
already enter into the scaling regime.
Encouraged by this finding,
we turn to the analysis of the 
inverse
correlation-length critical exponent $1/\nu$.

\begin{figure}
	\includegraphics[width=90mm]{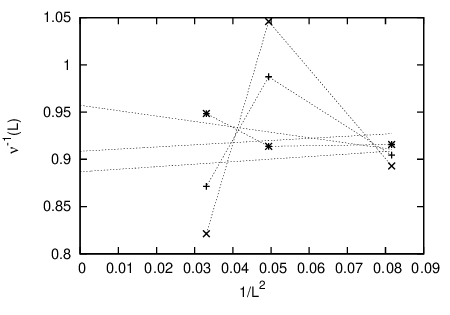}
\caption{\label{figure4}
The approximate inverse correlation-length critical exponent 
	$\nu^{-1}(L)$ (\ref{approximate_inverse_correlation_length_critical_exponent})
	is plotted for $1/L^2$ with $\Delta=0$.
The parameter $\delta$ is set to
	($+$) $\delta=0.5+3.96/6$ (\ref{delta_1.16}),
	($\times$) $1$,
	and
	($*$) $1.4$.
The least-squares fit to these data yields
	$1/\nu=0.91(14)$, $0.89(27)$, and
	$0.957(32)$,
	respectively, in the thermodynamic limit.
}
\end{figure}

In Fig. \ref{figure4},
as the symbol $(+)$,
we present the approximate inverse correlation-length critical exponent
$\nu^{-1}(L)$ for $1/L^2$ with
the same $\delta$ and $\Delta$ as those of Fig. \ref{figure2}.
Here, the approximate inverse correlation-length critical exponent is given by
\begin{equation}
\label{approximate_inverse_correlation_length_critical_exponent}
	\nu^{-1}(L-1/2)=\frac{\ln \chi_F^{(3)}(L)|_{K=K_c^*(L)} -\ln \chi_F^{(3)}(L-1)|_{K=K_c^*(L-1)}}{3(\ln L-\ln (L-1))}
	,
\end{equation}
because 
$\chi_F^{(3)}$'s peak develops as $\sim L^x$
with $\chi_F^{(3)}$'s scaling dimension $x=3/\nu$
(\ref{scaling_dimension}).
The least-squares fit to the ($+$) data yields an estimate $1/\nu=0.91(14)$
in the thermodynamic limit.
Additionally, 
similar analyses were made for the values of ($\times$) $\delta=1$,
and ($*$) $1.4$ with $\Delta=0$, and the results are shown
in Fig. \ref{figure4}.
The least-squares fit to these data yields
$1/\nu=0.89(27)$ and $0.957(32)$, respectively.
As shown in
Fig. \ref{figure4},
around 
such 
an
optimal regime of $\delta$,
the $\nu^{-1}(L)$ data
indicate
the value, $1/\nu \approx 0.9$,
and the deviation seems to be bounded by
\begin{equation}
\label{inverse_correlation_length_critical_exponent}
1/\nu=0.90(15)
.
\end{equation}
The validity of this estimate is cross-checked in Sec. \ref{section2_2},
where we make
the scaling plot for $\chi_F^{(3)}$
in order to examine its scaling behavior further in detail.
As mentioned above,
the $\nu^{-1}(L)$ data exhibit a rather suppressed finite-size artifact.
Such a feature was noticed by  the pioneering fidelity-susceptibility analysis
\cite{Yu09},
where the two-dimensional cluster with only $N \le 20$ spins 
yields reliable estimates for the critical exponents.

\begin{figure}
	\includegraphics[width=90mm]{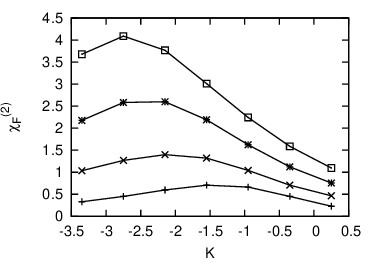}
\caption{\label{figure5}
	The ordinary second-order fidelity susceptibility $\chi^{(2)}_F$
	is plotted for various $K$, and 
	($+$) $L=3$,
	($\times$) $4$,
	($*$) $5$, and
	($\Box$) $6$;
	the parameters are the same as those of Fig. \ref{figure2}.
	There should exist non-analytic contributions because of its small scaling dimension
	$\approx 1.8(=0.9\cdot 2)<2$, Eq. (\ref{scaling_dimension}) and
	(\ref{inverse_correlation_length_critical_exponent}).
}
\end{figure}

As a comparison,
in Fig. \ref{figure5},
we show the ordinary second-order fidelity susceptibility 
$\chi^{(2)}_F(=-\partial^2_\lambda F|_{\lambda=0})$
(see Eq. (\ref{fidelity_susceptibility}));
here, the parameters are
the same as those of Fig. \ref{figure2}.
The scaling dimension of the ordinary fidelity susceptibility $\chi_F^{(2)}$ should be $\approx 1.8(=0.9\cdot 2)$
(see Eq. (\ref{scaling_dimension}) and (\ref{inverse_correlation_length_critical_exponent})), which does not reach 
extensive-quantity's dimensionality, $2$. Therefore, it does not fit detailed analysis of criticality.

We address a number of remarks.
First,
the result $1/\nu=0.90(15)$ [Eq. (\ref{inverse_correlation_length_critical_exponent})]
appears to be substantially smaller than
that of
the
3D-$XY$ universality class, $1/\nu(\approx 0.67^{-1})\approx 1.49$
\cite{Campostrini06,Burovski06}, for instance.
Therefore, it is suggested that the criticality belongs to a unique universality
class.
Last, the scaling dimension $x=3/\nu$
(\ref{scaling_dimension})
of the high-order fidelity susceptibility
is given solely by $\nu$,
and we do not have to make any additional 
considerations 
of the order parameters and the
associated scaling dimensions.
We stress that the perturbation operator $V$ 
(\ref{perturbation}) is not a symmetry breaking field,
and so,
the scaling theory
\cite{You11,Schwandt09,Albuquerque10} applies to the present treatment.
Hence,
the present analyses are all closed within the $\nu$-mediated singularities.

\subsection{\label{section2_2}
$XY$-VBS phase transition with $\Delta=0$ and fixed $\delta$:
Scaling plot of $\chi_F^{(3)}$}

In this section,
as a crosscheck,
we make the scaling plot for $\chi_F^{(3)}$,
relying on the analysis 
in Sec. \ref{section2_1}.
For that purpose,
we need to fix the scaling dimension $x$ of $\chi_F^{(3)}$.
Putting
the exponent $1/\nu=0.9$ [Eq. (\ref{inverse_correlation_length_critical_exponent})]
into the scaling relation
$x=3/\nu$ (\ref{scaling_dimension}),
we obtain
$\chi_F^{(3)}$'s scaling dimension as
\begin{equation}
\label{scaling_dimension2}
x=2.7
.
\end{equation}

\begin{figure}
	\includegraphics[width=90mm]{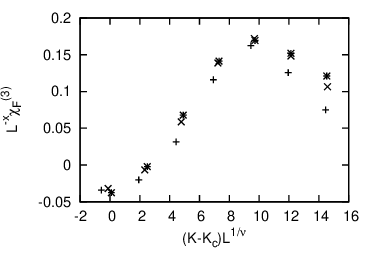}
\caption{\label{figure6}
The scaling plot,
	$(K-K_c)L^{1/\nu}$-$L^{-x}\chi_F^{(3)}$,
is presented for 
	($+$) $L=4$,
	($\times$) $5$, and
	($*$) $6$,
	based on the scaling formula (\ref{scaling_formula});
	the interaction parameters,
	$\delta=0.5+3.95/6$ (\ref{delta_1.16}),
	and $\Delta=0$,
	are the same as those of Fig. \ref{figure2}.
Here, the scaling parameters are set to
$K_c=-3.76$ (\ref{critical_point}),
$1/\nu=0.9$ (\ref{inverse_correlation_length_critical_exponent}), and
$x=2.7$ (\ref{scaling_dimension2}).
The scaled data collapse into a scaling curve satisfactorily.
}
\end{figure}

In Fig. \ref{figure6},
we present the scaling plot,
$(K-K_c)L^{1/\nu}$-$L^{-x}\chi_F^{(3)}$, for 
($+$) $L=4$,
($\times$) $5$, and
($*$) $6$.
Here, the parameters,
$\delta=0.5+3.95/6$ (\ref{delta_1.16})
and $\Delta=0$, are the same as those of Fig. \ref{figure2}.
The scaling parameters are set to
$K_c=-3.76$ [Eq. (\ref{critical_point})],
$1/\nu=0.9$ [Eq. (\ref{inverse_correlation_length_critical_exponent})],
and
$x=2.7$ [Eq. (\ref{scaling_dimension2})].
The scaled data seem to collapse into a scaling curve satisfactorily.
Particularly, the 
($\times$) $L=5$ and 
($*$) $6$ data are about to overlap each other.
Hence,
the validity of the scaling analysis in Sec. \ref{section2_1} is confirmed;
the corrections to finite-size scaling seems to be rather small,
as observed in Sec. \ref{section2_1}.
We stress that there is no {\it ad hoc} adjustable parameter
undertaken
in the scaling plot, Fig. \ref{figure6}.

This is a good position
to address an overview on the related studies of the critical exponent $\nu$.
To the best of author's knowledge,
little attention has been paid to the $XY$-VBS critical exponents
for the $XY$ (U$(1)$ symmetric) frustrated magnet.
On the one hand, as for the Heisenberg (SU$(2)$ symmetric) magnet,
the frustration-driven phase transition from the N\'eel phase
to the quantum disordered state
has been studied rather extensively \cite{Nishiyama12,Wang16,Nomura21,Liu22,Liu22b}.
As shown in Table \ref{table},
various frustrated magnets such as
the $J_1$-$J_2$-$Q$ \cite{Nishiyama12},
$J_1$-$J_2$ \cite{Wang16,Nomura21,Liu22},
and
$J_1$-$J_2$-$J_3$ \cite{Liu22b} Heisenberg models, have been simulated
with the exact diagonalization (ED),
tensor product states (TPSs),
restricted Boltzmann machine combined with pair product states (RBM$+$PP),
projected entangled pair state (PEPS),
and 
tensor network (TN) methods, respectively.
The estimated correlation-length critical exponents 
seem to be unsettled.
Almost all values are larger than that of the ordinary 3D-$XY$ universality class
$\nu \approx 0.67$
\cite{Campostrini06,Burovski06},
indicating that a novel class of phase transition may set in.
The present estimate $\nu=1.11(19)$ for the U$(1)$-symmetric case also 
exhibits a tendency for an enhancement, as compared to that of the 3D-$XY$ universality class.

\begin{table}
\caption{
	The present result $1/\nu=0.90(15)$ [Eq. (\ref{inverse_correlation_length_critical_exponent})] by means of the exact diagonalization (ED)
method is compared with a number of related simulation studies of the
	Heisenberg (SU(2) symmetric) models \cite{Nishiyama12,Wang16,Nomura21,Liu22,Liu22b}.
So far, a variety of technics such as the tensor product states (TPSs),
restricted Boltzmann machine combined with the
pair product states (RBM$+$PP),
projected entangled pair state (PEPS),
and 
tensor network (TN) methods
have been applied to the SU(2) frustrated magnets.
	}
\label{table}       
\begin{tabular}{llll}
\hline\noalign{\smallskip}
	model &  method & $\nu $  \\
\noalign{\smallskip}\hline\noalign{\smallskip}
	$J_1$-$J_2$-$K$ (this work) & ED & $1.11(19)[=0.90(15)^{-1}]$  \\
	$J_1$-$J_2$-$Q$ \cite{Nishiyama12} & ED& $1.1(3)$  \\ 
	$J_1$-$J_2$ \cite{Wang16} & TPSs & $0.50(8)$  \\ 
	$J_1$-$J_2$ \cite{Nomura21} &RBM$+$PP & $1.21(5)$  \\ 
	$J_1$-$J_2$ \cite{Liu22} &PEPS & $0.99(6)$  \\ 
	$J_1$-$J_2$-$J_3$ \cite{Liu22b} &TN & $\approx 1.0$  \\ 
\noalign{\smallskip}\hline
\end{tabular}
\end{table}

Last, we address a remark
on the conventional (second order)
fidelity susceptibility $\chi_F^{(2)}$.
Because the correlation-length critical exponent takes a large value, $\nu(=0.9^{-1})\approx 1.11$,
the scaling dimension of the conventional fidelity susceptibility, 
$x^{(2)}(=2/\nu)\approx 1.8$ \cite{You11,Schwandt09,Albuquerque10},
does not reach extensive-quantity's scaling $L^2$ in two spatial dimensions.
Therefore, the conventional fidelity susceptibility 
does not capture the underlying singularity properly.

\subsection{\label{section2_3}
Multi-criticality of the $XY$-VBS phase boundary: Crossover-scaling analysis of $\chi_F^{(3)}$}

In this section, we investigate the end-point singularity
of the $XY$-VBS phase boundary at the multi-critical point $J_2=0.5$.
For that purpose, we turn on the anisotropy parameter $\Delta=-0.15$ in this section; see Introduction.
We then resort to the crossover-scaling theory \cite{Riedel69,Pfeuty74},
where the distance from the multi-critical point $\delta$ is incorporated
into the scaling formula (\ref{scaling_formula}) accompanying the crossover critical exponent $\phi$.
The crossover-scaling formula is given by the expression
\begin{equation}
\label{crossover_scaling_formula}
\chi^{(3)}_F= L^{\dot{x}} g\left(
(K-K_c(\delta))L^{1/\dot{\nu}}  ,
\delta L^{\phi/\dot{\nu}} \right)
,
\end{equation}
with 
the phase-transition point $K_c(\delta)$ for $\delta$,
and
a scaling function $g$.
Here, the multi-critical exponents, $\dot{x}$ and $\dot{\nu}$,
denote $\chi_F^{(3)}$'s scaling dimension and correlation-length critical exponent, respectively,
right at $\delta=0$.
The crossover critical exponent $\phi$ describes the singular part of the phase boundary
\cite{Riedel69,Pfeuty74}
\begin{equation}
\label{phase_boundary}
K_c (\delta) \sim - \delta^{1/\phi}
,
\end{equation}
toward $\delta \to 0^+$ ($J_2\to 0.5^-$).

Before commencing the crossover-scaling analysis of $\chi_F^{(3)}$,
we fix the multi-critical exponents, $\phi$, $\dot{\nu}$, and $\dot{x}$.
We make a proposition that the phase boundary (see Eq. (\ref{phase_boundary})) terminates linearly
at $\delta=0$ ($J_2=0.5$)
\begin{equation}
\label{crossover_critical_exponent}
\phi=1
.
\end{equation}
Such a linearity implies that the phase boundary 
$K_c(\delta) \sim - \delta$ could be interpreted as a self-dual line \cite{Qin17,Jian17}.
In Ref. \cite{Qin17},
the duality between the VBS and easy-plane-N\'eel phases is argued
for the easy-plane-$J$-$Q$ and bose-Hubbard models.
As for the multi-critical correlation-length
critical exponent, we accept 
\begin{equation}
\label{multi-critical_correlation_length_critical_exponent}
\dot{\nu}=0.5
.
\end{equation}
This exponent was obtained by the above-mentioned duality analysis of the simulation results \cite{Qin17},
as well as
the hyperscaling relation $\dot{\alpha}=2-(d+z)\dot{\nu}$ 
with the spacial dimensionality $d=2$,
multi-critical specific-heat critical exponent $\dot{\alpha}=0$ (mean-field),
and the dynamical critical exponent $z=2$ \cite{Dutta98,Hornreich75,Diehl00,Diehl01}.
The hyperscaling relation holds generically \cite{Albuquerque10}, and 
it
was derived for the fully-frustrated magnets
(Lifshitz criticality) field-theoretically \cite{Dutta98,Hornreich75,Diehl00,Diehl01};
the above value $z=2$ is taken from this argument in the mean-field limit.
The value $z=2$ indicates that the imaginary-time correlation length $\xi_\tau$
diverges as $\xi_\tau \sim \xi^2$ ($\xi$: real-space correlation length).
The imaginary-time correlation length is substantially longer than the real-space one.
In the exact-diagonalization analysis, one does not have to care about this space-time anisotropy,
because at the ground state, the inverse temperature $\beta$ diverges as $\beta \to \infty$,
which covers $\xi_\tau(\ll \beta)$.
Therefore, in the following crossover-scaling analysis,
we are allowed to concentrate only on the real-space sector, $L$.
Lastly,
putting $\dot{\nu}=0.5$ 
[Eq. (\ref{multi-critical_correlation_length_critical_exponent})]
into the scaling relation (\ref{scaling_dimension}),
we obtain the multi-critical scaling dimension of $\chi_F^{(3)}$
\begin{equation}
\label{multi-critical_scaling_dimension2}
	\dot{x}=6
.
\end{equation}

\begin{figure}
	\includegraphics[width=90mm]{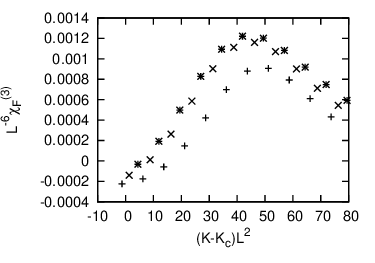}
\caption{\label{figure7}
The crossover scaling plot,
$(K-K_c(\delta))L^{1/\dot{\nu}}$-$L^{-\dot{x}}\chi_F^{(3)}$,
is presented for
($+$) $L=4$,
	($\times$) $5$, and
	($*$) $6$
	with $\Delta=-0.15$ and $\dot{x}=6$ (\ref{multi-critical_scaling_dimension2})
	The second argument of the crossover-scaling formula 
	(\ref{crossover_scaling_formula})
	is fixed to a constant value $\delta L^{\phi/\dot{\nu}}=25$
	with the crossover critical exponent $\phi=1$ (\ref{crossover_critical_exponent})
	and
	the multi-critical correlation-length critical exponent $\dot{\nu}=0.5$
	(\ref{multi-critical_correlation_length_critical_exponent}),
	and $K_c$ is determined via the same scheme as that of Sec. \ref{section2_1}.
	The present estimate $\phi=1$ implies that the phase boundary ends up
	linearly toward the multi-critical point as in Eq. (\ref{phase_boundary}).
}
\end{figure}

In Fig. \ref{figure7},
we present the crossover scaling plot, 
$(K-K_c(\delta))L^{1/\dot{\nu}}$-$L^{-\dot{x}}\chi_F^{(3)}$,
for various system sizes
($+$) $L=4$,
($\times$) $5$, and
($*$) $6$, with
the anisotropy parameter $\Delta=-0.15$ and the multi-critical
scaling dimension $\dot{x}=6$ (\ref{multi-critical_scaling_dimension2}).
The second 
argument of the crossover-scaling formula 
(\ref{crossover_scaling_formula}) is fixed to
$\delta L^{ \phi/\dot{\nu}}=25$ with 
$\phi=1$ (\ref{crossover_critical_exponent})
and 
$\dot{\nu}=0.5$ (\ref{multi-critical_correlation_length_critical_exponent}),
and the critical point $K_c$ is determined via the same scheme as that of Sec. \ref{section2_1}.
The crossover-scaled data appear to collapse into a scaling curve satisfactorily.
Particularly, the 
($\times$) $L=5$ and
($*$) $6$ data are about to overlap each other.

We address a number of remarks.
First, the crossover critical exponent $\phi=1$ is supported.
As mentioned above, this result indicates that the phase boundary is interpreted as 
a self-dual line.
In the above-mentioned $XY$-paramagnetic phase boundary \cite{Dutta98,Diehl00,Diehl01}
for the frustrated magnet,
the crossover critical exponent takes $\phi <1$, and the phase boundary should curve convexly
as in Eq. (\ref{phase_boundary}); in this case, the $XY$ and paramagnetic phases are not
dual, as anticipated.
Last, the exponent $1/\nu=0.9$ (\ref{inverse_correlation_length_critical_exponent})
in the scaling plot, Fig. \ref{figure6},
and the multi-critical exponent $1/\dot{\nu}=2$ (\ref{multi-critical_correlation_length_critical_exponent})
in the crossover-scaling plot,
Fig. \ref{figure7}, differ significantly.
Hence, the underlying singularities behind these scaling plots
are not identical, and the overlap of the crossover-scaling plot is by no means accidental.

\section{\label{section3}Summary and discussions}

The $J_1$-$J_2$-$K$ $XXZ$ model (\ref{Hamiltonian}) was 
investigated with
the exact diagonalization method.
The $XY$-VBS phase transition was analyzed with
the high-order fidelity susceptibility
$\chi_F^{(3)}$
(\ref{fidelity_susceptibility})
\cite{Wang09,Lv22}, which is readily evaluated via the exact diagonalization scheme.
As a preliminary survey, 
with $\Delta=0$ and $\delta=0.5+3.95/6$ (\ref{delta_1.16}) fixed,
we evaluated $\chi_F^{(3)}$, and obtained 
the critical point $(J_{2c},K_c)=(-0.032(70),-3.76(42))$ (\ref{critical_point}).
The estimate agrees with the stochastic-series-expansion  result 
$(J_{2c},K_c)=(0,-3.95)$ \cite{Sandvik02}.
Thereby, we analyzed 
the inverse correlation-length critical exponent $1/\nu$
via
the scaling behavior of $\chi_F^{(3)}$ with $\Delta=0$ and fixed $\delta$.
We
estimate
the exponent
as $1/\nu=0.9(15)$ [Eq. (\ref{inverse_correlation_length_critical_exponent})],
and made a comparison with
those obtained for
the frustrated Heisenberg magnets \cite{Nishiyama12,Wang16,Nomura21,Liu22,Liu22b}.
We then turn to the analysis of the crossover-criticality
toward $\delta \to 0^+$ ($J_2 \to 0.5^-$) under the setting $\Delta=-0.15$.
With crossover-scaling the $\delta$ parameter properly,
the $\chi_F^{(3)}$ data were cast into
the crossover-scaling formula
(\ref{crossover_scaling_formula})
under the
propositions, $\phi=1$ (\ref{crossover_critical_exponent})
and $\dot{\nu}=0.5$ (\ref{multi-critical_correlation_length_critical_exponent}).
Thereby, we found that the crossover-scaled data fall into a scaling curve satisfactorily.
The linearity of the phase boundary $\phi=1$
implies that the phase boundary is interpreted as a self-dual line,
as claimed for the easy-plane $J$-$Q$ magnet \cite{Qin17,Jian17}.

The scaling dimension $x=3/\nu$
(\ref{scaling_dimension})
of $\chi_F^{(3)}$
is given solely by $\nu$,
and 
the undertaken analyses are closed within the $\nu$-mediated singularities.
Rather intriguingly,
according to the duality theory \cite{Qin17},
the $\nu$ value is related to the anomalous dimensions of the order parameters beside
the phase boundary.
It is tempting to consider those order parameters and the associated anomalous dimensions.
This problem is left for the future study.

\section*{Acknowledgment}
This work was supported by a Grant-in-Aid
for Scientific Research (C)
from Japan Society for the Promotion of Science
(Grant No. 
20K03767).

\section*{Data Availability Statement}

	My manuscript has no associated data.
Data will be made available on reasonable request.



%



%
%




\end{document}